\documentclass[12pt, prd, showpacs]{revtex4}
\usepackage{amsmath}
\usepackage{amssymb}

\input{tcilatex}

\begin{document}

\title[Short title for running header]{Focusing versus defocusing properties
of truly naked black holes}
\author{Naresh Dadhich }
\affiliation{Inter-University Centre for Astronomy\&Astrophysics, Post Bag 4, Pune 4
11007, India}
\email{nkd@iucaa.ernet.in}
\author{Oleg B. Zaslavskii}
\affiliation{Astronomical Institute of Kharkov V.N. Karazin National University, 35
Sumskaya St., Kharkov, 61022, Ukraine}
\email{ozaslav@kharkov.ua}

\begin{abstract}
We study the properties of the congruence of null geodesics propagating near
the so-called truly naked horizons (TNH) - objects having finite Kretschmann
scalar but with diverging tidal acceleration for freely falling observers.
The expansion of outgoing rays near the future horizon always tends to
vanish for the non-extremal case but may be non-zero for the distorted
(ultra)extremal one. It tends to diverge for the ingoing ones if the the
null energy condition (NEC) is satisfied in the vicinity of the horizon
outside. However, it also tends to zero for NEC violating cases except the
remote horizons. We also discuss the validity of test particle approximation
for TNHs and find the sufficient condition for backreaction remaining small.
\end{abstract}

\maketitle




\section{Introduction}

In recent years the family of relativistic objects was extended. In
particular, there are objects which have standard black holes features but
also have the properties which are very distinct from them. This includes
so-called naked black holes (NBH) and truly naked black holes (TNBH)\ \cite%
{nk1} - \cite{4n} and quasi-black holes which do not share all properties
with such black holes but also reveal naked behavior (see \cite{qbh} and
references therein). For definition of these objects we shall adhere to \cite%
{vo}, \cite{tr}. Let the geometry of the black hole horizon be as usual
characterized by finiteness of the curvature components calculated in the
orthonormal frame. We will be dealing with static geometries only. In such a
case, the Kretschmann scalar, $\mathcal{K}$ represents the sum of squared
components (see details below in Sec. II), so that the necessary and
sufficient conditions of the finiteness of $\mathcal{K}$ $\ $reduce to the
finiteness of each term separately. For this reason, if $\mathcal{K}$ is
finite, any other curvature invariant is also finite, so that in what
follows we will speak about the finiteness of $\mathcal{K}$ that is quite
sufficient to ensure the absence of standard scalar curvature sanguinities.
Let $Z$ be the typical combination of curvature components responsible for
transverse tidal acceleration and $Z=0$ at the horizon in the static frames
(the precise definition and explanations will be given below). We denote $%
\bar{Z}$ its value in the freely falling frame. Then, the horizon is
characterized as follows: (i) usual if $\bar{Z}=0$, (ii) naked if $\bar{Z}%
\neq 0$ but finite and (iii) truly naked if $\bar{Z}$ is infinite. Strictly
speaking, the term "naked" can only be justified if $\bar{Z}$ can be made as
large as one likes by changing the parameters of configurations \cite{nk1}, 
\cite{nk2} but in what follows we will use it without making this
reservation. The most pronounced and unusual distinction between static and
freely falling frame manifests itself in TNBHs whose properties rely heavily
on the geometry of space-time near the horizon. For this reason, the
vicinity of the horizon in such metrics deserves special treatment. In
particular, in the previous paper \cite{4n} the breakdown of analyticity and
extendability of the metric across the horizon in Kruskal-like coordinates
was considered in detail.

Meanwhile, the question remains what happens to the geometry of such
systems. The behavior of null geodesics near the horizon is tightly bound to
the nature of horizon. That is, we wish to study in this paper propagation
of null geodesic congruence near the horizon and thereby characterize truly
naked horizons (TNHs).

Although the properties of TNBHs look quite unusual, they may appear in a
natural way both in Eninsetin and non-Enisntein theories of gravity. Thus,
some exact solutions of the Branse-Dicke theory exhibit the corresponding
behavior \cite{br1}, \cite{br2}. In general relativity, TNBHs may appear if
near the horizon the radial pressure $p_{r}$ and density $\rho $ are related
by the simplest equation of state $p_{r}=w\rho $, provided the parameter $w$
is constrained to some special interval (see Sec. VB of \cite{4n} for
details). There exists also the 2+1 analogue of TNBHs (see Sec. 8.2 of \cite%
{k}).

Apart from pure theoretical interest, some motivation for studying TNBH
comes from the discussion on interpretation of observational data. It was
argued in \cite{abr} that the most part of data which are usually considered
as confirming the existence of black holes can be ascribed to black hole
mimickers - the objects with a size slightly bigger than a gravitational
radius. In the recent work \cite{mc} it was pointed out that, for a wide
class of such mimickers, gravity forces grow unbound near the horizon. In
doing so, the key role in physical interpretation is played by the
observations of luminescent objects which become strongly deformed near the
horizon (see Sec. III A 3). As the gravity is anomaly strong also near the
truly naked horizons, TNBHs can, at least in principle, reveal themselves as
a third kind of objects that takes part in competition which kind of objects
fits observational data better. Correspondingly, study of propagation of
light near such objects can supply us with information which would enable us
to oppose TNBH both to usual black holes and black hole mimickers.

The paper is organized as follows. In next two Secs. II and III, we set up
the framework in terms of the metric, tidal acceleration, metric near
horizon and expansion of the null congruence. It is followed by the
discussion of focusing and defocusing effects by using the Raychaudhuri
equation in Sec IV. In Secs. V and VI, we consider the finiteness of
curvature invariants under perturbations and distorted TNHs respectively. We
conclude with discussion.

\section{Metric and tidal acceleration}

Consider the metric%
\begin{equation}
ds^{2}=-Udt^{2}+H^{-1}dr^{2}+r^{2}d\Omega ^{2}  \label{met}
\end{equation}%
supported by the stress-energy tensor, $T_{\mu }^{\nu }=diag(-\rho
,p_{r},p_{\perp },p_{\perp })$. For this metric, the non-vanishing curvature
components in the orthonormal frame are given by 
\begin{equation}
R_{0r}^{0r}=E=-\frac{H^{\prime }\Phi ^{\prime }}{2}-H(\Phi ^{\prime \prime
}+\Phi ^{\prime 2})\text{, }U\equiv \exp (2\Phi )\text{,}  \label{01}
\end{equation}%
\begin{equation}
R_{0\theta }^{0\theta }=\bar{E}=-\frac{H\Phi ^{\prime }}{r}\text{, }R_{\phi
\theta }^{\phi \theta }=F=\frac{1-H}{r^{2}}\text{,}  \label{02}
\end{equation}%
\begin{equation}
R_{r\theta }^{r\theta }=\bar{F}=-\frac{H^{\prime }}{2r}\text{,}  \label{12}
\end{equation}%
here primes denotes derivative with respect to $r$. The components in Eqs. (%
\ref{01}), (\ref{02}) are responsible for the geodesic deviation of
geodesics in the corresponding direction. For what follows, it is convenient
to introduce the quantity%
\begin{equation}
Z=\bar{F}-\bar{E}=4\pi (p_{r}+\rho )=\frac{1}{2}(G_{r}^{r}-G_{0}^{0})\text{.}
\label{zr}
\end{equation}%
In the freely-falling frame by integration of geodesics equation we obtain
(see Eq. (12) of \cite{tr}) 
\begin{equation}
\bar{Z}=2\varepsilon ^{2}Y-Z\text{, }  \label{y}
\end{equation}%
\begin{equation}
Y\equiv \frac{Z}{U}\text{,}  \label{yz}
\end{equation}%
$\varepsilon $ being the energy per unit mass. The comoving energy density
measured by a freely falling radial observer%
\begin{equation}
\bar{\rho}=\frac{\varepsilon ^{2}Y}{4\pi }-p_{r}\text{.}
\end{equation}

We assume that in the vicinity of the horizon the metric behaves like%
\begin{equation}
U\sim (r-r_{h})^{q}\text{, }H\sim (r-r_{h})^{p}\text{, \thinspace }
\label{pq}
\end{equation}%
where $p>0$ and $q>0$. Then, we have near the horizon 
\begin{equation}
Z\sim (q-p)(r-r_{h})^{p-1}\text{.}  \label{z}
\end{equation}%
\begin{equation}
\bar{Z}\sim (q-p)(r-r_{h})^{p-q-1}\text{.}  \label{zz}
\end{equation}

Thus $Z\rightarrow 0$ always but the situation with $\bar{Z}$ may be
different and that's what leads to the classification of horizons. It is the
quantity $Y$ which leads to the possibility of essential enhancement of
curvature components. The fact that separate curvature components can be
enhanced greatly or even could be made infinite is reconciled with the
finiteness of the Kretschmann scalar $\mathcal{K=R_{\alpha \beta \gamma
\delta }R^{\alpha \beta \gamma \delta }}$ in terms of the Lorentz boost and
signature of the metric. In the static frame $\mathcal{K}$ $=4E^{2}+8\bar{E}%
^{2}+8\bar{F}^{2}+4F^{2}$, so that all terms enter $\mathcal{K}$ with the
"+" sign and its finiteness requires finiteness of each term separately.
However, in the freely falling frame there are mixed components which enter
the sum with different signs and cancel each other so that the overall sum
remains finite even when some components may be infinite.

\section{Metric near horizon and expansion}

\subsection{Metric near horizon}

We will be interested in the behavior of geodesics near the horizon.
However, the original coordinates $t,r$ become ill-defined there. It is
therefore desirable to rewrite the metric in the Kruskal-like coordinates.
To this end, we use the basic framework of Ref. \cite{4n} and write it in
the form 
\begin{equation}
ds^{2}=-dt^{2}A+\frac{du^{2}}{A}+r^{2}d\Omega ^{2}\equiv
ds_{2}^{2}+r^{2}d\Omega ^{2}  \label{metu}
\end{equation}%
where $A$ and $r$ are functions of a new radial quasi-global coordinate $u$
and $d\Omega ^{2}=d\theta ^{2}+\sin ^{2}\theta d\phi ^{2}$. Let there be a
horizon at $r=r_{h}$. The coordinate $u$ will in general have the finite
value $u_{h}$ on the Killing horizon \cite{br1}, \cite{br2}, \cite{kb01},
and hence there is no loss of generality in setting $u_{h}=0$. However,
there also exists in principle the case of infinite $u_{h}$ which
corresponds to so-called remote horizon when space-time on one side of the
"horizon" is geodesically complete. We will also point out such cases.

In terms of the quasi-global coordinates, the metric reads

\begin{equation}
A(u)=u^{n}F(u)\text{, }F(0)<\infty \text{, }n=const\text{.}  \label{au}
\end{equation}

Then, comparing Eqs (\ref{met}) and (\ref{metu}), we obtain%
\begin{equation}
r-r_{h}\sim u^{s}\text{, }s=\frac{2}{q-p+2}\text{, }n=\frac{2q}{q+2-p}.
\label{ru}
\end{equation}

We take $q>p-2$ whereas $q\leq p-2$ indicates to remote horizons. Let us
further introduce the tortoise-like coordinate $x$%
\begin{equation}
dx=\frac{du}{A}
\end{equation}%
and 
\begin{equation}
V=t+x\text{, }W=t-x.
\end{equation}%
The next step is to introduce Kruskal like coordinates $v$ and $w$: $V=V(v)$%
, $W=W(w)$, so we write 
\begin{equation}
ds_{2}^{2}=-AdWdV=-2Bdwdv  \label{b}
\end{equation}%
where the coefficient $B=\frac{A}{2}\frac{\partial W}{\partial w}\frac{%
\partial V}{\partial v}$ is finite and non-zero on the horizon. On the
horizon $v=const,$ or $w=const$ and it is convenient to put these constants
to zero. Then, the past horizon is given by 
\begin{equation}
V\rightarrow -\infty \text{, }W\text{ fixed}\longleftrightarrow v=0\text{, }w%
\text{ fixed,}
\end{equation}%
and the future horizon by 
\begin{equation}
W\rightarrow \infty \text{, }V\text{ fixed, }\longleftrightarrow w=0\text{, }%
v\text{ fixed.}
\end{equation}%
In what follows we shall only refer to the future horizon. It is clear that
for non-zero and finite $B$ near the horizon we require $\frac{dW}{dw}%
=const.A^{-1}$. Further, let us write 
\begin{equation}
r-r_{h}=f(u)\text{, }u=\chi (w,v)
\end{equation}%
where $f(u)=const.u^{s}$ as $u\rightarrow 0$. Again, near the horizon $u\sim
w$ \cite{br1, br2, kb01}, and we have in this region 
\begin{equation}
\chi (w,v)=wa(v)
\end{equation}%
where $a(v)\neq 0$. As a result, we obtain 
\begin{equation}
\frac{\partial r}{\partial v}\sim u^{s}  \label{drv}
\end{equation}%
and%
\begin{equation}
\frac{\partial r}{\partial w}\sim u^{s-1}  \label{drw}
\end{equation}%
near the future horizon. Below, we will exploit these useful relations.

\subsection{Expansion}

We consider propagation of radial null geodesics for the metric under
discussion. Let $\left( l^{\pm }\right) ^{\mu }$ be the tangent to such
geodesics for the metric (\ref{b}), $l^{\mu }l_{\mu }=0$ where $"+/-"$ refer
to outgoing/incoming geodesics. We write $l^{+}=B^{-1}\partial _{v}$, $%
l^{-}=B^{-1}\partial _{w}$, then the expansion $\theta \equiv l_{;\mu }^{\mu
}$ is given by 
\begin{equation}
\theta ^{out}=\frac{2}{rB}\frac{\partial r}{\partial v}
\end{equation}%
for outgoing geodesics and 
\begin{equation}
\theta ^{in}=\frac{2}{rB}\frac{\partial r}{\partial w}
\end{equation}%
for incoming ones. Then, Eqs (\ref{drv}) and (\ref{drw}) lead to 
\begin{equation}
\theta ^{out}\sim u^{s}\sim r-r_{h}\rightarrow 0\text{.}  \label{yin}
\end{equation}%
Thus, for the outgoing rays we obtain the standard result that the expansion
goes to zero at the horizon. However, for the incoming rays propagating
towards the future horizon, the situation is not so definite and it depends
on the parameter $s$. From Eqs (\ref{ru}) and (\ref{drw}), we write 
\begin{equation}
\theta ^{in}\sim -u^{s-1}\sim -n^{d}\text{, }s-1=\frac{p-q}{q+2-p}\text{, }d=%
\frac{2(p-1)}{2-p},  \label{in}
\end{equation}%
where $n$ is the proper distance. Note that for incoming geodesics $\theta
^{in}<0$ because $W$ grows with decreasing $r$ ($\frac{dx}{du}>0$ while $%
\frac{dr}{du}>0$ and $\frac{\partial x}{\partial W}=-\frac{1}{2}<0$, $\frac{%
\partial w}{\partial W}>0$, so that $\frac{\partial r}{\partial W}<0$).

\section{Raychaudhuri equation}

It is instructive to trace the features under discussion in the Raychaudhuri
equation. In general, it has the form 
\begin{equation}
\frac{d\theta }{d\lambda }=-\frac{\theta ^{2}}{2}-\sigma _{\alpha \beta
}\sigma ^{\alpha \beta }+\omega _{\alpha \beta }\omega ^{\alpha \beta
}-R_{\alpha \beta }l^{\alpha }l^{\beta }  \label{r0}
\end{equation}%
where $\lambda $ is the canonical parameter, $\sigma _{\alpha \beta }$ is
the shear tensor, $\omega _{\alpha \beta }$ is the rotation tensor. It is
clear that the space-time under study is free of shear and rotation, so the
equation reduces to 
\begin{equation}
\frac{d\theta }{d\lambda }=-\frac{\theta ^{2}}{2}-2\omega ^{2}Y\text{, }
\label{r}
\end{equation}%
where $Y$ is as defined in (\ref{y}) and the frequency%
\begin{equation}
\omega \equiv -l_{0}=\left( \frac{\partial v}{\partial t}\right) _{x}=\frac{%
\partial v}{\partial V}
\end{equation}%
is constant along the incoming ray $V=const.$. Although there is no comoving
frame for a null particle, the definition of $\bar{Z}$ implies simply the
limiting transition to the speed of light for a moving particle. For
incoming geodesics the affine parameter near the horizon is related to the
coordinate $u$ as 
\begin{equation}
d\lambda =-Cdu  \label{u}
\end{equation}%
where $C>0$ is some constant (see Sec. III of \cite{4n}). Using Eqs (\ref{zz}%
) , (\ref{ru}), (\ref{in}) and (\ref{r}) it is easy to see that for each
pair $(p,q)$ the asymptotic behavior Eq. (\ref{in}) agrees with Eq (\ref{r}).

We shall collect together various cases in a table below. As is known, the
energy condition plays the determining role in the Raychaudhuri equation and
we shall therefore indicate the status of the null energy condition (NEC) in
the vicinity of the horizon in each case ($+,0,-$ will indicate $\rho
+p_{r}>0,=0,<0$). On the horizon itself $\rho +p_{r}=0$ as usual.

\skip 0.5cm $%
\begin{tabular}{|l|l|l|l|l|l|}
\hline
&  & Type of BH horizon & $\theta ^{in}$ & NEC & $Y$ on the horizon \\ \hline
1 & $p=q=1$ & usual or naked & $<0$ & $0$ & finite \\ \hline
2 & $1<p<\frac{3}{2}$, $q=2-p$ & truly naked & $0$ & $-$ & $-\infty $ \\ 
\hline
3 & $p=\frac{3}{2}$, $q=\frac{1}{2}$ & naked & $0$ & $-$ & $<0$ \\ \hline
4 & $\frac{3}{2}<p<2$, $q=2-p$ & usual & $0$ & $-$ & $0$ \\ \hline
5 & $p\geq 2$, $p<q$ & truly naked & $-\infty $ & $+$ & $\infty $ \\ \hline
6 & $p\geq 2$, $p=q\geq 2$ & usual or naked & $<0$ & $0$ & finite \\ \hline
7 & $p\geq 2$, $q<p<q+1$ & truly naked & $0$ & $-$ & $-\infty $ \\ \hline
8 & $p\geq 2$, $q=p-1$ & naked & $0$ & $-$ & $<0$ \\ \hline
9 & $p\geq 2$, $q+1<p<q+2$ & usual & $0$ & $-$ & $0$ \\ \hline
10 & $p\geq 2$, $p\geq q+2$ & remote & $-\infty $ & $-$ & $0$ \\ \hline
\end{tabular}%
$

\skip 0.5cm

Now, some comments on the results presented in Table 1 are in order. Cases 1
and 6 mean that $\rho +p_{r}=0$ in the vicinity of the horizon in the main
approximation since in Eq. (\ref{z}) the factor $q-p$ vanishes. Then, the
knowledge of numbers $p$ and $q$ is insufficient and the behavior of $%
p_{r}+\rho $ depends on further details of the asymptotic form of the
metric. Correspondingly, in these cases the main terms of the order $%
(r-r_{h})^{-1}$ in $\bar{Z}$ in Eq. (\ref{zz}) also cancel, so near the
horizon $\bar{Z}$ has the order $(r-r_{h})^{0}$, its sign depends on the
further details of the system. In case 5, the horizon itself realizes the
focusing theorem (see, e.g., Sec. 2. 4.5 of textbook \cite{poisson})
according to which the geodesies are focused during the evolution of the
congruence (that indicates the occurrence of a caustic), provided NEC is
satisfied. In doing so, the canonical parameter of the geodesics remains
finite in view of Eq. (\ref{u}) and the location of caustic coincides with
the horizon.

It is clear that TNH can occur either for $p<q$ (Case 5) provided NEC is
satisfied near the horizon or for $p>q$ (Cases 2 and 7) with the violation
of NEC there. Note that $q$ and $p$ respectively indicate approach of $%
-g_{tt}=U$ and $g^{rr}=H$ in the metric (\ref{met}) to zero near the
horizon. The former refers to the Newtonian potential while the latter to
gravitational self interaction (field energy) \cite{naresh}. When $q>p$, $A$
would approach zero faster than $B$ indicating dominance of the field energy
over the Newtonian potential near the horizon. Here the NEC is satisfied.
For $p>q$, the opposite is the case, dominance of the Newtonian potential
over field energy which leads to the violation of NEC.

If space-time is asymptotically flat at infinity with NEC marginally ($\rho
+p_{r}=0$) satisfied, then $\theta ^{in}\rightarrow 0$ there. In cases 2-4
and 7-10, $\theta ^{in}=0$ at the horizon. Since $\theta ^{in}<0$ between
the horizon and infinity (at least in some vicinity of the horizon), it
cannot have a monotonic behavior. This means that the null congruence
suffers defocusing in some region which may be in the vicinity of the
horizon where the phantom matter density dominates over the focusing effect
of the second term in the Raychaudhuri equation (\ref{r}). Thus, only from
boundary conditions (at infinity and the horizon) we can extract some useful
information about the behavior of null congruence beam in the intermediate
region.

In case 3 the horizon is simple in that $A\sim u$, $\frac{dr}{du}$ is
finite, so the metric coefficients are analytical functions of $u$ and the
metric can be extended across the horizon. As null geodesics cross the
horizon with $\theta ^{in}=0$ and $\frac{d\theta ^{in}}{d\lambda }>0$, the
expansion becomes positive beyond the horizon. Thus, the defocusing effect
prevails at least in immediate vicinity of the horizon. We would like to
note that case 3 cannot be realized with the linear equation of state $%
p_{r}=w\rho $ near the horizon ($w<-1$ for phantom matter) because this
configuration has been shown to be singular \cite{p2}. The case 10 is
exceptional and it represents remote horizons where $\theta ^{in}\rightarrow
-\infty $ when $\lambda \rightarrow \infty $.

In general, we see that horizon itself can have defocusing effect on the
congruence in that $\theta ^{in}$ can vanish on the horizon and change its
sign (if analytical continuation is possible). But $\theta ^{in}$ cannot
become positive near the horizon if space-time has the usual topology (with
area growing away from horizon). In this sense, a possible focusing effect
of the horizon is as expected much stronger than a possible defocusing one.

\section{Perturbations and finiteness of curvature invariants}

So far, we have considered propagation of null geodesics (motion of photons)
in the fixed background of a TNBH. The fact that tidal forces in the freely
falling frame can grow unbundedly near the horizon makes the question of
validity of test particle approximation non-trivial. It was observed in \cite%
{nk2} for NBHs that by adding even perturbation with small comoving density
one can gain large curvature invariants with test particle approximation
remaining valid. This issue becomes even more sharp for TNBH where tidal
forces not only become large but, by definition, infinite. To clarify this
issue, we follow on the lines of \cite{nk2}. In doing so, we do not need to
consider details of motion of shells in the given background but need only
to check whether the curvature invariants remain finite in spite of infinite
tidal acceleration in the freely falling frame. That means TNH can occur but
not usual naked curvature singularity (the latter would imply divergencies
of the Kretschmann scalar).

For definiteness and simplicity, we restrict ourselves to the
spherically-symmetric case and consider adding usual dust to the original
source as was done in \cite{nk2}. The total stress-energy tensor is given by 
\begin{equation}
\bar{T}_{\mu \nu }=T_{\mu \nu }+\rho u_{\mu }u_{\nu }
\end{equation}%
where $\rho $ is the comoving density and $u^{\mu }$ is the four-velocity.
Then, the invariant $R_{\mu \nu }R^{\mu \nu }\sim \bar{T}_{\mu \nu }\bar{T}%
^{\mu \nu }$ would read as 
\begin{equation}
\bar{T}_{\mu \nu }\bar{T}^{\mu \nu }=T_{\mu \nu }T^{\mu \nu }+2\rho T_{\mu
\nu }u^{\mu \nu }+\rho ^{2}\text{.}  \label{tt}
\end{equation}

The second term is proportional to $\rho \bar{Z}$ which is given in Eq. (\ref%
{zz}). The rough estimate for $\rho $ can be written as

\begin{equation}
\rho \sim \frac{M_{p}}{4\pi r^{2}l}
\end{equation}

where $l$ is the proper size of dust distribution, $M_{p}$ is the proper
mass of dust. Near the horizon with the metric coefficient given by Eq. (\ref%
{pq}), $l\sim (r-r_{h})^{1-\frac{p}{2}}$. Then, the second term in (\ref{tt}%
) is of the order $(r-r_{h})^{c}$ where $c=\frac{3}{2}p-q-2$. For $c>0$ the
correction is negligible and it is finite for $c=0$. The former condition
reads

\begin{equation}
p>\frac{2}{3}(q+2)\text{.}  \label{p>q}
\end{equation}

This condition only ensures finiteness of curvature invariants. Table 1
lists only three cases, 2 ($1<p<\frac{3}{2}$, $q=2-p$), 5 ($p\geq 2,p<q$)
and 7 ($q<p<q+1$) for TNBHs \cite{tr}, \cite{4n}. The first one has
contradiction with the above inequality (\ref{p>q}) while the other two
respectively require $q>4$ and $q>1$.

All this only implies sufficient condition under which perturbation does not
lead to formation of naked singularities with diverging curvature
invariants. Whether it is necessary or not is not quite clear. Because of
strong non-linearity of the field equation, one should solve dynamic
situation in a self-consistent manner. This is beyond the scope of the
present paper, we have only addressed the question for the test field
approximation.

\section{Null geodesics near distorted truly naked horizons}

Let us generalize the framework by relaxing the requirement of spherical
symmetry. This not only leads to increasing the number of possibilities but
also gives rise to a new qualitative feature which was absent in spherical
symmetry.

Consider the generic static metric which can be written in the form%
\begin{equation}
ds^{2}=-N^{2}dt^{2}+dn^{2}+\gamma _{ab}dx^{a}dx^{b}
\end{equation}%
where $a,b=1,2$. Again, we consider for definiteness the future horizon. Let
us concentrate on the behavior of null generators of the horizon which can
be obtained as the limit of outgoing rays. We want to examine what happens
to the expansion of the horizon%
\begin{equation}
\theta ^{out}=\frac{\partial _{\mu }(N\sqrt{\gamma }l^{\mu })}{\sqrt{\gamma }%
N}  \label{dexp}
\end{equation}%
where $l^{\mu }=(\Omega ,\Gamma N,l^{a})$ is the tangent to the geodesic.
Note that $\Omega $ remains finite near the horizon.

As is known, $\theta ^{out}=0$ if the horizon is usual and non-extremal. In
the present consideration, neither of these properties hold good apriori.
This means we should check whether or not $\theta ^{out}=0$ on the horizon
afresh. We call the horizon extremal if $N\sim \exp (-\frac{n}{n_{0}})$ with 
$n\rightarrow \infty $ and ultraextremal if $N\sim n^{-m}$ with $m>0$. The
examples are the Reissner-Nordstr\"{o}m metric with a charge equal to mass
(the extremal case) and Reissner-Nordstr\"{o}m-de Sitter (the ulatraextremal
case) with a special relation between charge, mass and cosmological constant 
\cite{romans}.

What would still hold good would be that it is a null geodesic ($g_{\mu \nu
}l^{\mu }l^{\nu }=0,a^{\mu }=l_{;\nu }^{\mu }l^{\nu }=0$) and it tends to
null generator ($l_{\mu }\sim \left( N^{2}\right) _{\mu }$ as $N\rightarrow
0 $) of the horizon.

We give the explicit components of acceleration, 
\begin{equation}
a^{1}=(l_{,0}^{1}+NN^{\prime }l^{0})l^{0}+l_{,1}^{1}l^{1}+(l_{,a}^{1}-\frac{1%
}{2}\frac{\partial \gamma _{ab}}{\partial n}l^{b})l^{a}\text{,}  \label{a1}
\end{equation}%
\begin{equation}
a^{0}=l_{,0}^{0}l^{0}+l_{,1}^{0}l^{1}+2\frac{N^{\prime }}{N}l^{0}l^{1}+\frac{%
N_{,a}}{N}l^{a}l^{0}  \label{a0}
\end{equation}%
\begin{equation}
a^{a}=[l_{,0}^{a}+NN_{;b}\gamma
^{ab}l^{0}]l^{0}+l_{,1}^{q}l^{1}+(l_{,b}^{a}+\Gamma _{bc}^{a}l^{c})l^{b}%
\text{,}  \label{aa}
\end{equation}
which would be needed later.

\subsection{Non-extremal case}

Near the horizon 
\begin{equation}
N=\kappa n+\kappa _{3}(x^{a})n^{3}+...\text{, }n\rightarrow 0\text{,}
\end{equation}%
$\kappa =const\neq 0$ is the surface gravity. This relation is insensitive
to whether the horizon is usual, naked or truly naked, it only requires
finiteness of the Kretschmann scalar \cite{vis}. Then, $\frac{l_{a}}{l_{1}}%
\sim \frac{N_{,a}}{N^{\prime }}$, so that near the horizon $%
l_{a}=c_{a}(x^{b})n^{3}$ where $c_{a}$ are finite. Neglecting higher orders
in $n$, $\Gamma \approx \Omega $ near the horizon, it is easy to check that
the component $a^{b}\rightarrow 0$ automatically while the other components
in the leading order read as 
\begin{equation}
a^{1}\approx \Omega _{h}N(2\Omega ^{\prime }N^{\prime }+\Omega
_{,0})_{h}+O(N^{2}),
\end{equation}%
\begin{equation}
a^{0}=\Omega _{h}(2\Omega ^{\prime }N^{\prime }+\Omega _{,0})_{h}+O(N),
\end{equation}%
where subscript "h" means that the quantity is calculated on the horizon.

It is worth noting that, although the metric is static, the components of
the vector $l^{\mu }$ depend in general on time that can be checked, for
example, in the spherically symmetric case (see Sec. III))

It follows from the geodetic character of the null generator of horizon that 
\begin{equation}
(2\Omega ^{\prime }N^{\prime }+\Omega _{,0})_{h}=0.  \label{con}
\end{equation}

By substitution in (\ref{dexp}), we find that the contributions from $%
\partial _{0}$ and $\partial _{n}$ mutually cancel out due to (\ref{con})
while contribution of $\partial _{a}$ is of higher order $n^{2}$. Thus we
establish the expected result that $\theta ^{out}=0$ for all non-extremal
horizons, may it be usual, naked or truly naked.

\subsection{Extremal and ultraextremal cases}

Let the metric coefficients near the horizon have the form (extremal case)%
\begin{equation}
N=B_{1}(x^{a})\exp (-\frac{n}{n_{0}})+B_{2}(x^{a})\exp (-\frac{2n}{n_{0}}%
)+O(\exp (-\frac{3n}{n_{0}}))\text{, }n_{0}=const\text{, }n\rightarrow
\infty \text{,}
\end{equation}%
\begin{equation}
\gamma _{ab}=\gamma _{ab}^{(H)}+\gamma _{ab}^{(1)}\exp (-\frac{n}{n_{0}}%
)+O((\exp (-\frac{2n}{n_{0}})).
\end{equation}

TNBH\ is realized if $B_{1}$ or $B_{2}$ depend on $x^{a}$ \cite{vo,tr}.
Then, we can write $\frac{\partial N}{\partial n}\sim N\sim N_{,a}\sim l_{a}$
and writing $l_{a}=c_{a}N$ where $c_{a}\neq 0$ is finite on the horizon, it
follows that in the horizon limit the term proportional to $\left(
B_{1}\right) _{,a}c_{,b}\gamma ^{ab}$ persists and does not in general
vanish. As a result, for a TNBH $\theta ^{out}\neq 0$ on the horizon in
general. Meanwhile, for usual extremal horizon, $B_{1}=const$, so that $%
\theta ^{out}\rightarrow 0$.

In the ultraextremal case we have 
\begin{equation}
N=\frac{A_{1}(x^{a})}{n^{m}}+\frac{A_{2}(x^{q})}{n^{m+1}}+O(n^{-m-2})\text{, 
}m>0\text{.}
\end{equation}%
Then, on the same lines it follows that in general $\theta ^{out}$ will not
be zero on horizon because now $A_{1}(x^{a})$ is not constant on the
horizon. The same argument will be true for shear $\sigma _{ab}$ which may
as well not vanish on the horizon.

As far as the behavior of ingoing rays is concerned, there would be increase
in number of cases in Table 1 due to deviation from spherical symmetry
indicated by $B_{i}$, $A_{i}$ being function of $x^{a}$. It was however
already observed earlier \cite{vo} that extremal or ulrtraextremal TNH may
have properties very different from those of the usual horizons. For
example, the structure of the Einstein tensor (and, correspondingly, the
structure of the stress-energy tensor) on the horizon is quite different for
the two cases \cite{vis}. The case of non-vanishing expansion on the horizon
is another unusual property of ultraextremal TNHs.

\section{Discussion}

Through the study of behavior of null geodesics in vicinity of horizon, we
would like to point out some universal features of spherically symmetric
horizons. Instead of tidal forces $Z$ (or $\bar{Z}$) used above to
distinguish different kinds of horizons, we can speak in terms of the null
energy density, $\rho _{n}=\rho +p_{r}$ which is equivalent to $Z$ according
to (\ref{zr}). Then, the finiteness of the Kretschamnn scalar requires $\rho
_{n}=0$ on the horizon (see, e.g., \cite{vis}) irrespective of the kind of
the horizon. The distinction depends on properties of the boosted quantity $%
\bar{\rho}_{n}$ measured by a freely falling observer. We note the
following: (a) whenever $\bar{\rho}_{n}=0$, the horizon is usual, finite $%
\bar{\rho}_{n}\neq 0$ corresponds to a naked horizon and infinite $\bar{\rho}%
_{n}$ represents a TNH. (b) As expected, the expansion of outgoing null
congruence always vanishes at the non-extremal horizon irrespective of its
character. (c) What is rather unexpected is the possibility of non-vanishing
expansion of the outgoing congruence for extremal or ultraextremal horizons.
(d) For the incoming light rays both focusing and defocusing effects are
possible. The expansion is negative and finite for the usual/naked horizon
with $\rho _{n}=0$ while it always vanishes (except for remote horizon when
it is negative and infinite) for $\rho _{n}<0$ in the vicinity of the
horizon, and it is negative and infinite for $\rho _{n}>0$. More exactly,
the focusing is severest with $\theta ^{in}=-\infty $ on horizon for $\rho
_{n}>0$ and also for the remote horizon even when $\rho _{n}<0$ in some
vicinity of the horizon while in all other cases it is defocusing for $\rho
_{n}<0$ with $\theta ^{in}$ approaching zero on horizon. This means that
defocusing caused by presence of negative $\rho _{n}$ in vicinity of horizon
is just able to counteract the inherent focusing as horizon is approached.
While positive $\rho _{n}$ further strengthens focusing and it reaches
divergent proportion on horizon.

We would like to stress that TNBHs not only represent pure theoretical new
objects with interesting and unusual properties but they could in reality be
realized as configurations with negative pressure and scalar fields as they
in fact arise naturally in the context of (2+1) gravity \cite{4n}. We have
here considered only static TNBHs but the really interesting and pertinent
question is, how could they be formed in the course of gravitational
collapse of a realistic matter distribution?

\begin{acknowledgments}
O.Z. thanks the Inter-University Centre for Astronomy and Astrophysics
(IUCAA), where the most part of this work was performed, for hospitality and
stimulating atmosphere.
\end{acknowledgments}

\end{document}